# A novel approach for structure analysis of two-dimensional membrane protein crystals using x-ray powder diffraction data

R. A. Dilanian[1*], C. Darmanin[2], J. N. Varghese[2], S. W. Wilkins[3], T. Oka[4,5], N. Yagi[5], H. M. Quiney[1] and K. A. Nugent[1]

[1]ARC Centre of Excellence for Coherent X-ray Science, School of Physics, The University of Melbourne, VIC 3010, Australia

[2]ARC Centre of Excellence for Coherent X-ray Science, CSIRO Preventative Health Flagship, CSIRO Molecular Health and Technologies, Melbourne, VIC 3052, Australia

[3]ARC Centre of Excellence for Coherent X-ray Science, CSIRO Material Science and Engineering, Melbourne, VIC 3168, Australia

[4]Department of Physics, Faculty of Science and Technology, Shizuoka University, Shizuoka, 422-8529, Japan

[5]Research and Utilization Division, SPring-8/JASRI, Hyogo 679-5198, Japan

[*]E-mail: roubend@unimelb.edu.au

## Abstract

The application of powder diffraction methods in two-dimensional crystallography is regarded as intractable because of the uncertainties associated with overlapping reflections. Here, we report an approach that resolves these ambiguities and provides reliable low-resolution phase information directly from powder diffraction data. We apply our method to the recovery of the structure of the bacteriorhodopsin (bR) molecule to a resolution of 7Å using only powder diffraction data obtained from two-dimensional purple membrane (PM) crystals.

# I. INTRODUCTION

The majority of known membrane proteins are much more likely to form two-dimensional (2D), rather than three-dimensional (3D) crystals during the crystallization processes [1] which limits the possibility of obtaining their molecular structures using the standard methods of protein crystallography. Powder diffraction methods are, in contrast, not critically sensitive to the quality and dimensions of crystals and this suggests their use in the structure analysis of 2D crystals [2, 3]. It seems to be widely believed, however, that powder diffraction methods are not suitable for high-resolution structure analysis of large proteins. The present letter challenges this view by performing the structure determination of a membrane protein using powder diffraction data.

The main impediment to the use the powder diffraction data in structural analysis of 2D crystals is the frequent occurrence of completely overlapping reflections. Of the 17 possible crystallographic plane groups, 8 describe unit cells with equal cell parameters ($a = b$) [4]. For such symmetries, all $(h,k)$ and $(k,h)$ reflections are completely overlapped. For a 2D crystal with average unit cell dimensions of 60 Å, such completely overlapping reflections form approximately 50% of the total number of diffraction peaks up to 7 Å resolution. This statistic apparently renders structure analysis from powder diffraction data practically impossible.

# II. RESULTS AND DISCUSSION

We here consider a cluster of atoms defined by the irreducible representation of the atomic positions in the unit cell, as the constituent element of the crystal structure. The cluster can include several, single, or even part of the membrane proteins. It contains no internal symmetry operation except the unitary operator. For convenience of the further analysis we will call this cluster a molecule. In this representation, the diffraction pattern of the 2D crystal can be regarded as the superposition of the scattering from the individual molecules, the molecular form factor (MFF), and the Laue interference function (see Methods section). It has been shown [5] that the powder diffraction pattern exhibits a sensitivity to the one-electron density of the molecule. In the case of the low-resolution imaging of a scattering object of characteristic dimension, $D$, scattering from a molecule comprised of discrete atoms is almost indistinguishable from that of an homogeneous and continuous envelope distribution of electron density if the criterion $SD \leq 1$ is satisfied.

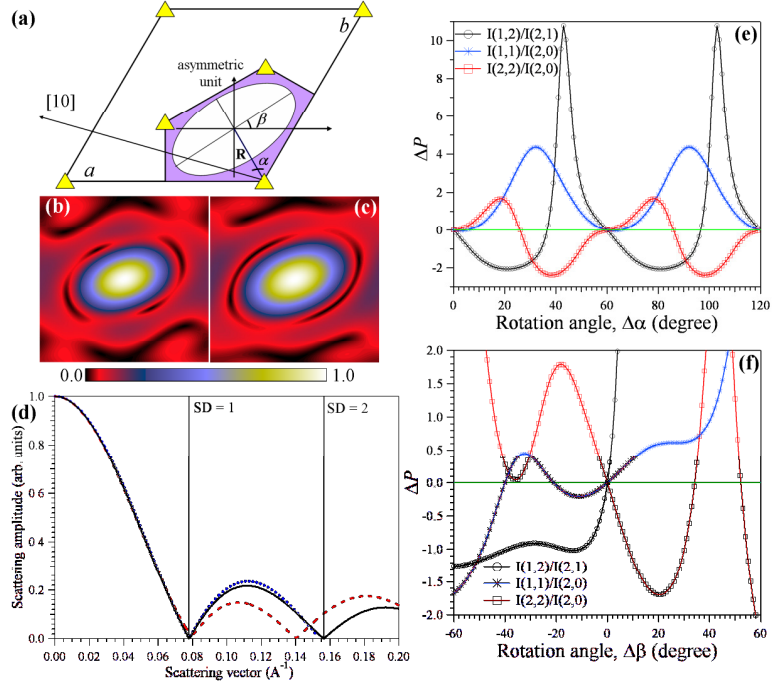

**FIG. 1**. (a) The model of the molecular envelope located in the asymmetric unit of the *p3* symmetry unit cell. The low-resolution shape and size of the envelope is matched with the form and size of the asymmetric unit. The parameters $R$, $\alpha$, and $\beta$ define the origin and orientation of the molecule in the unit cell. (b) & (c) The normalized MFF calculated for 7 Å resolution using discrete[6] and continuous representation of bR, respectively. (d) The variation of the MFF of bR along the [1,0] direction calculated for different resolutions: 7 Å (dotted curve), 5 Å (dashed curve). The solid curve represents the case of the continuous representation of bR. (e) & (f) Intensity ratios of three pairs of reflections are represented as functions of (e) $\Delta\alpha = \alpha - \alpha_{True}$ (e) and (f) $\Delta\beta = \beta - \beta_{True}$. $\Delta P = P_C - P_{True}$ represents difference between the calculated values of intensity ratios and the exact values.[6] For each pair of reflections there are four possible values of $\Delta\alpha$ when $\Delta P = 0$. Only $\Delta\alpha = 0^0$ (the initial orientation), $\Delta\alpha = 60^0$ (the inversion operation), and $\Delta\alpha = 120^0$ (the 3-fold symmetry operation) correspond to such an orientation of the molecular envelope when $\Delta P = 0$ for all pair of reflections. Consequently, $\Delta P = 0$ for all pair of reflections only when $\Delta\beta = 0^0$.

The low-resolution shape of the molecule can be predicted by looking at the form of the asymmetric unit. Assuming that the size of the bR molecule is comparable with the size of the asymmetric unit, we can consider the continuous representation of the molecular envelope to be a solid ellipsoid, covering the maximum possible area of the asymmetric unit without intersecting its boundaries; this last requirement is to avoid overlapping of molecules under the appropriate symmetry operations, Fig. 1(a). The 7 Å resolution maps of MFF calculated using discrete [6] and continuous representation of the bR molecule are shown in Figs. 1(b) and (c), respectively. Fig. 1(d) shows the variation of the MFF of the bR molecule [6] along the [1,0] crystallographic direction calculated for different resolution limits. The solid curve represents the case of scattering from the molecular envelope assuming a uniform electron density distribution. It can be seen (Fig. 1(d)) that the MFF calculated at different resolutions are almost identical if the criterion $SD \leq 1$ is satisfied.

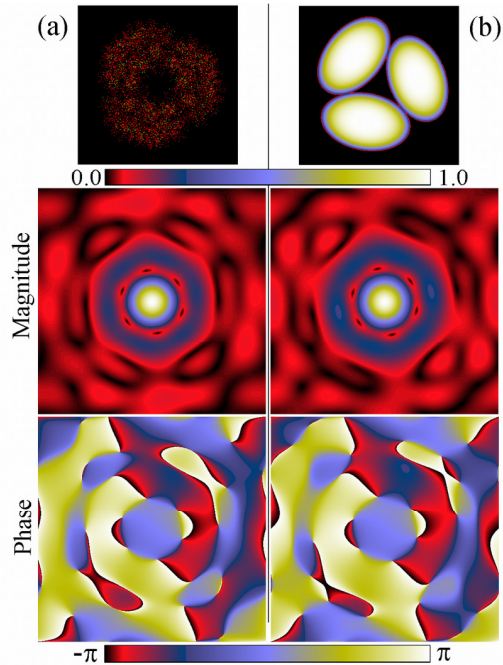

**FIG. 2**. The normalized molecular form factor of PM (a) simulated using discrete representation of the BR molecule, and (b) reconstructed using continuous representation of the BR molecule (b). The complex molecular form factor, $\Phi(\mathbf{S}) = A_\Phi(\mathbf{S})\exp(i\varphi_\Phi(\mathbf{S}))$, where $A_\Phi(\mathbf{S})$ is the magnitude and $\varphi_\Phi(\mathbf{S})$ is the phase of $\Phi(\mathbf{S})$, is described by Eq. 4 (Methods section).

We here define a two-dimensional molecular envelope function, $E(R,\alpha,\beta)$, describing the effective molecular boundaries. We require that the total density inside the envelope is equal to the total number of electrons in bR, and that $E(R,\alpha,\beta) \to 0$ elsewhere. The electron density is distributed inside the envelope in such a way that it is a maximum at the center of the molecular envelope and monotonically decreasing towards its boundary. In this way, different positions and orientations of this envelope will give different ratios of intensities for the pair of Bragg reflections, Fig. 1(e,f) and Fig. 4 (Methods section). For low-resolution structure analysis, therefore, the orientation and origin of the molecule, rather than the actual distribution of electron density inside the molecular envelope, is the dominant factor affecting the amplitude of the structure factor.

**TAB. 1.** The quality of the decomposition of overlapping reflections in comparison with the actual data (Model) [6]. The algorithm allows a highly accurate determination of the values of intensity ratios of completely overlapping reflections.

| Intensity ratios of completely overlapping reflections | | | |
|---|---|---|---|
| $(h_1,k_1)$ | $(h_2,k_2)$ | $P = I(h_1,k_1)/I(h_2,k_2)$ | |
| | | Reconstructed | Model |
| (1, 2) | (2, 1) | 1.61 | 2.11 |
| (1, 3) | (3, 1) | 0.73 | 0.77 |
| (2, 3) | (3, 2) | 0.60 | 0.47 |
| (1, 4) | (4, 1) | 1.96 | 1.95 |
| (2, 4) | (4, 2) | 1.61 | 1.48 |
| (1, 5) | (5, 1) | 3.00 | 7.47 |
| (3, 4) | (4, 3) | 0.16 | 0.15 |
| (2, 5) | (5, 2) | 0.40 | 0.98 |
| (1, 6) | (6, 1) | 0.25 | 0.02 |
| (2, 6) | (6, 2) | 0.96 | 1.16 |
| (1, 7) | (7, 1) | 0.95 | 0.59 |

Consider a pair of completely overlapping reflections, $I(h_m, k_m)$ and $I(k_n, h_n)$. The ratio $P_m = I(h_m, k_m)/I(k_n, h_n)$ can, therefore, be estimated by varying the origin and the orientation of the approximated molecular envelope and using intensity ratios of resolved reflections $P_{pq} = I(h_p, k_p)/I(h_q, k_q)$ as constraints (see Methods section). If one can measure $L$ independent ratios, which we label by the index $j = 1, \ldots, L$, then the mean discrepancy between the calculated and the measured intensity ratios for these resolved reflections can be calculated using

$$\Omega(R, \alpha, \beta) = \frac{1}{L} \sum_{j=1}^{L} \frac{\left(P_j^{calc}(R, \alpha, \beta) - P_j\right)^2}{\sigma_j^2}, \quad (1)$$

where $P_j$ and $P_j^{calc}(R, \alpha, \beta)$ are the $j^{th}$ measured and calculated intensity ratios, respectively, and $\sigma_j$ is the standard deviation of the measurements, $P_j$.

The ability of the method to decompose completely overlapping reflections was first tested with the diffraction data simulated using the results presented in Ref. [6]. The initial shape, size and the location of the molecular envelope were set as shown in Fig. 1(b). The result of the analysis is shown in Fig. 2 and in Tables 1 and 2. The algorithm allows the determination of the values of intensity ratios of completely overlapping reflections with a high level of accuracy. Fig. 2 shows the reconstructed and simulated MFF of PM. As one can see, the magnitudes and phases of calculated and simulated MFF are in good agreement. It can also be seen from Table 2 that the reconstructed and the simulated phases of all reflections have identical signs and comparable values.

We then analysed the powder diffraction pattern obtained from PM crystals [2]. The integrated intensities of diffraction peaks up to a 4 Å resolution were estimated using the Le Bail analysis [7], shown in Fig. 3(a) and Table 3. The initial set of reconstruction parameters were similar to that used for the simulated data. The completely overlapping reflections up to 7 Å resolution were decomposed using the method described previously. Structure factors were then analysed by the maximum entropy method [8] to obtain the 2D electron density map of PM. The resulting low-resolution 2D electron density map of PM is shown in Fig. 3(b). The map clearly indicates the envelope of a bR molecule and the electron density of the transmembrane α-helices.

**TAB. 2**. Structure factors reconstructed from simulated powder diffraction data of the PM crystal.

| (h, k) | Structure Factors | | | |
|---|---|---|---|---|
| | Amplitude | | Phase | |
| | Calculated | Model | Calculated | Model |
| (1, 1) | 11.91743 | 12.16437 | -22.91344 | -22.24714 |
| (2, 0) | 4.9046 | 4.85244 | 24.6692 | 23.02843 |
| (1, 2) | 1.24374 | 1.22668 | -44.89789 | -43.30173 |
| (2, 1) | 0.77099 | 0.5802 | 89.28897 | -88.30343 |
| (3, 0) | 0.54015 | 0.51799 | -18.73582 | -18.15258 |
| (2, 2) | 1.1924 | 1.27496 | -55.86939 | -56.06551 |
| (1, 3) | 0.80354 | 0.78751 | -70.60166 | -71.02457 |
| (3, 1) | 1.0992 | 1.02924 | -12.98028 | -15.74641 |
| (4, 0) | 1.04355 | 0.77478 | -69.13335 | -70.16235 |
| (2, 3) | 0.77105 | 0.78369 | -52.75801 | -45.45028 |
| (3, 2) | 1.28032 | 1.65788 | -16.40019 | -11.4538 |
| (1, 4) | 0.81414 | 0.87439 | -46.21956 | -33.86207 |
| (4, 1) | 0.41454 | 0.44938 | -33.43213 | -15.79022 |
| (5, 0) | 0.69898 | 0.61851 | -8.17534 | -16.61341 |
| (3, 3) | 0.12166 | 0.26063 | -70.46205 | -70.96063 |
| (2, 4) | 0.64201 | 0.62005 | 27.57895 | 26.81057 |
| (4, 2) | 0.39847 | 0.41971 | -56.18278 | -54.5848 |
| (1, 5) | 0.49246 | 0.41146 | -69.36095 | -69.64992 |
| (5, 1) | 0.16436 | 0.05511 | 18.04615 | 24.21409 |
| (6, 0) | 0.83752 | 0.55592 | -4.49881 | -12.19641 |
| (3, 4) | 0.16659 | 0.14832 | -72.33902 | -43.89246 |
| (4, 3) | 1.03 | 0.96659 | -50.48938 | -54.99652 |
| (2, 5) | 0.15321 | 0.40799 | 46.40117 | 48.11394 |
| (5, 2) | 0.3813 | 0.41697 | -59.58168 | -51.89029 |
| (1, 6) | 0.05669 | 0.00509 | 40.63563 | -20.25437 |
| (6, 1) | 0.23052 | 0.23092 | -20.9682 | -34.25779 |
| (4, 4) | 0.02474 | 0.06532 | -83.19175 | -49.10226 |
| (3, 5) | 0.6139 | 0.75898 | -81.36501 | -84.01214 |
| (5, 3) | 0.24176 | 0.15246 | 9.8293 | 18.73371 |
| (7, 0) | 0.03193 | 0.04496 | 46.43535 | 53.22326 |
| (2, 6) | 0.12654 | 0.11023 | 38.04956 | 44.08322 |
| (6, 2) | 0.13258 | 0.09543 | 54.00176 | 69.88409 |
| (1, 7) | 0.23082 | 0.18419 | -40.55927 | -44.48675 |
| (7, 1) | 0.24426 | 0.3144 | 42.18927 | 57.4036 |

The three helices located in the inner part of the bR molecule have projected density higher than the helices at the outer part of bR, which is in an agreement with results obtained previously [2, 6].

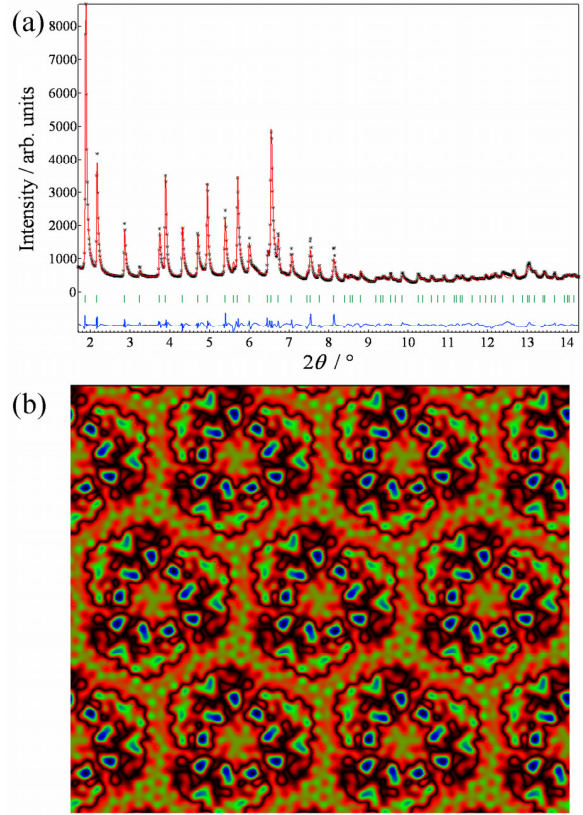

**FIG. 3**. (a) Results of the Le Bail analysis of the powder diffraction pattern obtained from PM crystals. The upper solid line (red) represents the calculated intensities and small asterisks (black) superimposed on it are the observed intensities. The lower solid line (blue) is the difference between the observed and calculated intensities. The short vertical lines (green) below the profiles indicate peak positions of possible reflections. (b) The resulting low-resolution 2D electron density map of PM reconstructed from experimental data.

## III. METHODS

### III.A MOLECULAR FORM FACTOR

The relation between an intensity of the diffraction peak and the crystal structure is,

$$I(\mathbf{S}) = \Lambda(\mathbf{S}) L_P(\mathbf{S}) |F(\mathbf{S})|^2, \qquad (2)$$

where $\mathbf{S}$ is the scattering vector, $\Lambda(\mathbf{S})$ is the Laue interference function, $L_P(\mathbf{S})$ is the Lorentz and polarization factor, and $F(\mathbf{S})$ is the structure factor.

**Table 3**. Unit-cell dimensions and *R* factors for the PM crystal. The Le Bail analysis was performed using RIETAN-2000 [9], and MEM analysis – using PRIMA[8]. The background of the powder diffraction pattern was represented by a composite background function obtained with PowderX [10]. The split pseudo-Voigt function of Toraya [11] was used as a profile function. The unit cell parameters of PM were determined using DICVOL [12].

| | |
|---|---|
| Plane Group | *p3* |
| Number of molecules in the unit cell | 3 |
| *a* (Å) | 61.081(1) |
| *b* (Å) | = *a* |
| $\gamma$ (°) | 120.0 |
| Data range, $2\theta$ (°) | 1.69–14.30 |
| Wavelength (Å) | 1.0 |
| Number of exp. points | 1110 |
| Number of reflection | 105 |
| Resolution (Å) | 4.0 |
| $R_{wp}$, % | 6.95 |
| $R_p$, % | 5.12 |
| $R_B$, % | 0.96 |
| $R_E$, % | 3.75 |
| $\gamma = R_{wp}/R_E$ (goodness-of-fit) | 1.85 |

If we consider an individual molecule as a constituent element of the crystal structure and assume that the asymmetric unit contains only one molecule then the structure factor can be written as,

$$F(\mathbf{S}) = \sum_{k=1}^{M} \Phi_k(\mathbf{S}) \cdot \exp\{i2\pi(\mathbf{S}\hat{G}_k \mathbf{R}_0)\}, \qquad (3)$$

where *M* is the number of elements of the symmetry group, $\hat{G}_k$ is the matrix representation of the $k^{\text{th}}$ symmetry element of the symmetry group, $\mathbf{R}_0$ is the origin of the molecule in the asymmetric unit, $\Phi(\mathbf{S})$ defines the scattering from the individual molecule or the molecular form factor (MFF),

$$\Phi_k(\mathbf{S}) = \sum_{j=1}^{N} f_j(S) \cdot \exp\{i2\pi(\mathbf{S}\hat{G}_k \Delta \mathbf{R}_j)\} T_j(\mathbf{S}), \quad (4)$$

where $S = |\mathbf{S}|$, $N$ is the number of atoms in the asymmetric unit, $f_j(S)$ is the atomic scattering factor of the $j^{th}$ atom, $\Delta \mathbf{R}_j = \mathbf{R}_j - \mathbf{R}_0$ is the relative position of the $j^{th}$ atom in the asymmetric unit, and $T_j(\mathbf{S})$ is a thermal distribution function which accounts for the motion of the $j^{th}$ atom. If the lattice displacements are independent and the atoms oscillate in a harmonic potential then $T_j(\mathbf{S})$ can be represented by a scalar term, $B_j(\mathbf{S})$, the Debye-Waller factor.

### III.B THE UNIQUENESS OF THE SOLUTION

If we consider the shape of the molecular envelope as an ellipsoid and assume that the electron density is uniformly distributed within the boundaries of the ellipsoid, then for the low-resolution limit the amplitude of the structure factor, $F(\mathbf{S})$, of a particular crystallographic plane $(h,k)$, is proportional to the length, $D$, of the central chord of the molecular envelope, which is parallel to the $(h,k)$ plane. The intensity ratio of two peaks, $I(h_1,k_1)$ and $I(h_2,k_2)$ can, therefore, be estimated as

$$P = I(h_1,k_1)/I(h_2,k_2) = (D_1/D_2)^2, \quad (5)$$

where $D_1 = 2\sqrt{a^2 \cos^2(\gamma) + b^2 \sin^2(\gamma)}$, $D_2 = 2\sqrt{a^2 \cos^2(\gamma+\varepsilon) + b^2 \sin^2(\gamma+\varepsilon)}$, $a$ and $b$ are main axes of the ellipsoid, $\gamma$ is the angle between the $a$ and the $(h_1,k_1)$ plane, and $\varepsilon$ is the angle between $(h_1,k_1)$ and $(h_2,k_2)$ planes, Fig. 4(a). Then the orientation of the envelope, $\gamma$, can be calculated as a function of the following parameters, $P$, $b/a$, and $\varepsilon$. Fig. 4(b) shows the variation of $\gamma$ as a function of $P$ for two diffraction peaks of PM crystal, $(1,0)$ and $(1,1)$. The vertical lines in Fig. 4(b) correspond to the intensity ratios 0.5 and 2.0. As one can see, there are four possible solutions, marked A, B, C, and D, corresponding to the selected values of the intensity ratio of two individual reflections. Pairs of solutions (A&B) and (C&D) are separated by the angle $\Delta\alpha = 60^0$, and related to each other through the inversion operation. Existence of (A&B) or (C&D) pairs has a simple geometrical interpretation. Due to the symmetry of the envelope, 2*mm*, two different orientations of the envelope gives exactly the same ratios of the central chords and, therefore, equal intensity ratios of the pair of reflections. We may call these differently oriented envelopes the crystallographic twins.

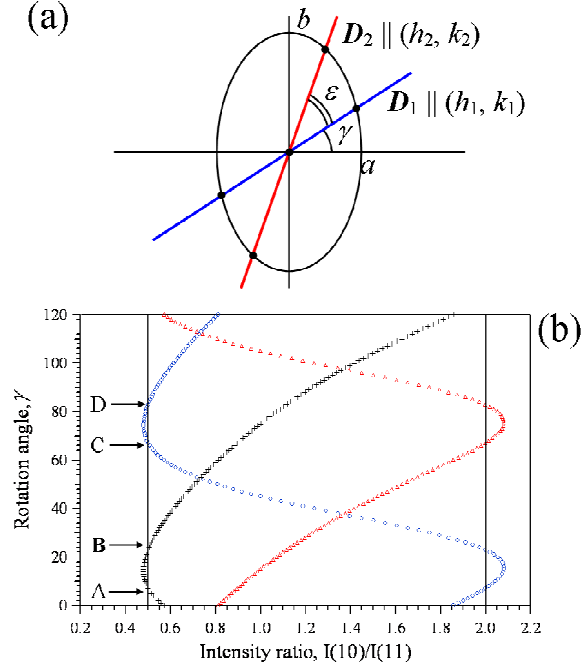

**FIG. 4** (a) The geometrical interpretation of Eq. (5). (b) The variation of the $\gamma$ parameter as a function of the intensity ratio for two diffraction peaks. A, B, C, and D indicate four possible solutions for $\gamma$, corresponding to the selected values of the intensity ratio of two individual reflections.

### III.C ALGORITHM

To decompose completely overlapping reflections, we devised the following iterative procedure, where indices *p* indicate a resolved reflection and *m* an overlapped reflection:

1. Intensity data are analyzed by the Le Bail method to obtain integrated intensities of resolved, $I(h_p, k_p)$, and overlapping, $I(h_m, k_m)$, reflections and to create a set of intensity ratios of resolved reflections, $P_j$.

2. The initial $E(R,\alpha,\beta)$ function is estimated using the initial crystallographic information.
3. $E(R,\alpha,\beta)$ is used to calculate intensities of all resolved reflections, $I(h_p,k_p)$.
4. The set of intensity ratios, $P_j^{calc}$, is calculated using results from Step 3.
5. Intensity ratios of all pairs of completely overlapping reflections are evaluated by minimizing the discrepancy function, Eq. (1). Since the total intensity in the overlapped diffraction ring, $I(h_m,k_m)+I(k_m,h_m)$, is known from experimental data, the individual components of the completely overlapping reflections can be determined.
6. The individual components of the first pair of overlapping reflections, $I(h_m,k_m)$ and $I(k_m,h_m)$ are added to the set of resolved reflections, $I(h_p,k_p)$. Steps 2-5 are repeated to decompose the remaining elements of the overlapped reflections. In such iterative processes, the number of resolved reflections will increase and the number of overlapping reflections will decrease after each loop in the calculation. This iterative approach improves the stability of the analysis.
7. The structure factors derived from Step 6 are analyzed by the Maximum Entropy Method to yield a low–resolution representation of the 2D electron density, $\rho(x,y)$. The final structure factors, $F_{MEM}(\mathbf{S})$, are calculated by the Fourier transform of $\rho(x,y)$ in the unit cell.

## IV. CONCLUSION

The approach described here has been applied to two-dimensional powder diffraction data. The extension to three-dimensional powder diffraction data is straightforward and will allow the development of an excellent starting point for the analysis of such data. This approach can, therefore, be used as a tool for preliminary structure analysis of sub–micron or nanoscale protein crystals with sizes that do not satisfy the requirements of single crystal structure analysis methods.

Our method avoids the complexity of single crystal structure analysis caused by the poor quality and the smallness of the crystals and allows the correct estimation of the integrated intensities and initial phases of reflections. Moreover, the combination of the single crystal and powder diffraction methods greatly enhances the ability of the structure analysis techniques and, therefore, will significantly expand the list of biological molecules the structures of which may be determined from X-ray diffraction data. We note that the availability of very bright X-ray free electron laser sources should allow the acquisition of very high−resolution powder diffraction data along with lower resolution single crystal data. The method here might therefore find application for data from these sources using weakly scattering biological nanoscale crystals.

**Acknowledgments**

The authors acknowledge the support of the Australian Research Council through its Centres of Excellence and Federation Fellowship programs, and from the CSIRO Preventative Health Flagship, Neurodegenerative Diseases Theme and the CSIRO, Emerging Science Initiative.

RAD is responsible for the bulk of the analysis, the computer programming and the writing. CD contributed to the data analysis and the writing. TO and NY contributed the experimental data and input on this. JNV, SWW, HMQ & KAN contributed to the formulation of the problem, the analysis and the writing of the paper.

None of the authors have a conflict of interest associated with this work.